# Investigation of Accelerated Hydration in Nanosilica Incorporated Tricalcium Silicate using THz Spectroscopy

Shaumik Ray[1,3], Jyotirmayee Dash[1,3], Nirmala Devi[1,3], Saptarshi Sasmal[2,3], Bala Pesala[1,3]

**Abstract** Incorporation of nanosilica in cementitious systems helps in accelerating the cement hydration process and early-stage formation of the main hydration product, i.e., calcium silicate hydrate (C-S-H). Nanosilica reacts with the other key hydration product, calcium hydroxide ($Ca(OH)_2$) to form additional C-S-H. In this work, Terahertz spectroscopy has been employed to study the variation in hydration kinetics of the main constituent of cement, Tricalcium Silicate (C3S), due to the addition of different percentages of nanosilica. The variation has been analyzed in terms of change in the absorbance as well as shift in the wavenumber of the key resonances. The resonance around 520/514 $cm^{-1}$, due to symmetric/asymmetric bending of $SiO_4$ tetrahedral units, decreases in intensity very rapidly as hydration progresses and more so for the nanosilica incorporated samples demonstrating accelerated hydration. Interestingly, the resonance around 451 $cm^{-1}$, due to symmetric and asymmetric bending of $SiO_4$ tetrahedral units along with some contribution of vibrational modes of CaO polyhedral units, does not show a similar decrease in intensity. The prominence of the resonance around 451 $cm^{-1}$ even after the onset of hydration is found to be mostly due to the contribution from the vibrational modes of the newly formed $SiO_4$ tetrahedra chains of C-S-H. Early-stage formation of C-S-H due to nanosilica incorporation has been analyzed by tracking the shift in resonances as polymerized chains of $SiO_4$ are formed. These results have been corroborated with Scanning Electron Microscopy results which clearly show the early changes in the morphology of the hydrated C3S due to nanosilica incorporation.

Electronic mail: balapesala@gmail.com

[1]*Council of Scientific and Industrial Research (CSIR), Central Electronics Engineering Research Institute (CEERI), Chennai - 600113, Tamil Nadu, India*

[2]*Council of Scientific and Industrial Research (CSIR), Structural Engineering Research Centre (SERC), Chennai - 600113, Tamil Nadu, India*

[3]*Academy of Scientific and Innovative Research, CSIR – SERC, Chennai - 600113, Tamil Nadu, India*

## 1 Introduction

Addition of nanomaterials in cementitious systems has been of great interest for developing a new class of cement-based building materials with improved properties and extraordinary functionalities. The incorporation of the nanomaterials not only accelerates the hydration rates but also results in an improved mechanical strength, fracture toughness, ductility of the cement composite/concrete along with enhanced durability. Tricalcium silicate (C3S), the most prevalent constituent (approximately 50-70%) in the cement matrix, is responsible for the early stage hydration as well as strength development in concrete [1]. C3S starts to react with water in the first few hours of the hydration process with the formation of two main hydration products - amorphous or poorly crystalline calcium silicate hydrate (C-S-H) and crystalline calcium hydroxide (($Ca(OH)_2$, CH). C-S-H is the most dominant hydration product since it accounts for around 70 – 80% of the hydration products formed during the process and is responsible for the strength development in concrete. On the other hand, CH is a so-called undesirable hydration product because of its leaching properties. Reducing the amount of cement (by incorporation of nanosilica) for the development of high strength concrete with faster setting times will have significant effect on reducing the carbon footprint as cement production is one of the most contributing sources for $CO_2$ emission. Recent surveys have mentioned that cement plants are responsible for 5-7% of the total carbon dioxide emissions globally [2]. Studies have shown that adding nanosilica has two effects on the hydration process of cementitious systems. Firstly, it fills the capillary voids in the hydrated (hydrating) cement resulting in an increased density which leads to greater mechanical strength of the concrete [3]. Secondly, nanosilica also reacts with the hydration product CH, which helps in the formation of extra C-S-H. The morphology and nanostructure of the C-S-$H_{II}$ formed due to the reaction between nanosilica and CH may differ from the C-S-$H_I$ formed during hydration of C3S only as shown in the equations below [4]:

$$\text{C3S} + H_2O \rightarrow \text{C-S-}H_I + Ca(OH)_2 \quad (1)$$

$$SiO_2 + Ca(OH)_2 \rightarrow \text{C-S-}H_{II} \quad (2)$$

The C-S-H gel has a complex structure that has a layer of calcium oxides in the center sandwiched by parallel silicate ($SiO_4$) chains in such a way that the oxygen atoms in the central layer are shared between the calcium atoms as well as the silicate chains [4]. Studies of the nanostructure of C-S-H shows that the structure resembles the pattern similar to crystalline calcium silicate hydrates such as 1.4 Tobermorite and Jennite [5]. Even though there is no confirmed structure that exactly resembles the C-S-H formed from hydration of cement, Tobermorite polymorphs and Jennite are considered to be model structures similar to C-S-H. Even though the nanostructure of the crystalline C-S-H like structures are similar, the basic difference between Tobermorite and Jennite is that, in Jennite, only half of the oxygen atoms in the central layer are shared with the $SiO_4$ chains, while the remaining oxygen atoms form a part of -OH groups [6]. This effect is seen to result in a significant corrugation of the Ca-O layers in Jennite. Some of the oxygen atoms in this layer are shared with the Dreierketten, (a chain of dimeric silicates tetrahedra, linked by a bridging silicate) while the others form a part of the water molecules and -OH groups. The -OH groups are balanced by calcium atoms, resulting in the formation of Ca-OH bonds. Contrary to what was believed earlier, no Si-OH bonds are present in a well-crystallized Jennite [7].

To study the properties of cementitious materials, conventional techniques such as X-ray diffraction (XRD), Isothermal Calorimetry, Nuclear Magnetic Resonance (NMR) spectroscopy, Scanning Electron Microscopy (SEM) and Transmission Electron Microscopy (TEM) are generally employed to understand the variation in phases of cement and cementitious systems during hydration [8]–[12]. Earlier studies using calorimetric measurements, thermogravimetric analysis (TGA), SEM and mercury intrusion porosimetry have shown different aspects of hydration of C3S due to nanosilica addition [3]. This study shows that the nanosilica accelerates C3S hydration till 3 days after which the rate is seen to recede. Studies on the optimization of the dosage of silica nanoparticles in C3S using XRD and TGA analysis have compared early stage (before 8 hours) effect of nanosilica incorporated samples in comparison to the control sample (sample without nanosilica) [13]. By day 1, 25% additional C-S-H is formed with the simultaneous reduction of 32% CH in the case of the sample with 10% nanosilica, which confirms the pozzolanic effect during hydration. Furthermore, FTIR spectroscopy (in the Mid-IR region) in this work also showed that the nanoparticle incorporation accelerates the polymerization of silicates and leads to the formation of more crystalline structures, which possibly indicates the formation of Tobermorite like C-S-H structures. SEM analysis carried out in this work confirmed the formation of a compact and dense microstructure by the first day of hydration. This densification of the microstructure also reduces the rate of hydration at a later stage. The accelerating effect of C3S hydration due to the addition of colloidal silica has been extensively analyzed using FTIR spectroscopy and Differential Scanning Calorimetry [14]. This study used MIR spectroscopy to analyze the hydration process of cement in terms of the variation in peak intensity as well as a shift in key resonances. Reduction in intensity of these peaks indicates the onset of reactions whereas the shift of these resonances to higher or lower wavenumbers implies polymerization or depolymerization effects during the formation of C-S-H. This paper has discussed the rate of hydration of only C3S (sample without nanosilica) and the C3S samples added with 2% and 5% colloidal silica. The spectra relating to pure C3S (817 – 953 $cm^{-1}$) due to anti-symmetrical stretching vibrations of $SiO_4$ tetrahedrons showed a faster dissolution rate for increasing colloidal silica percentage. In this study it has been pointed out that colloidal silica is not dissolved in the hydration process, but its surface provides highly reactive sites for the condensation of $SiO_4$ monomers released from C3S. The critical observation made in this paper is the broadening of the band around 1030 $cm^{-1}$ (due to Si-O stretching vibrations) as the amount of colloidal silica is increased. Furthermore, the peaks broaden in order to include vibrations of higher frequencies which is consistent with increased connectivity in the silicate network, implying a greater degree of polymerization. However, studies have shown that the degree of polymerization is lower at higher C/S ratios [6]. This points to the fact that the degree of polymerization of Tobermorite-like C-S-H structures (Ca/Si ratio of 0.83) is more than Jennite-like C-S-H structures (Ca/Si ratio of 1.5). FTIR spectroscopy has also been employed to explore the effect of temperature and the addition of 10% nanosilica on the C-S-H nanostructure due to C3S hydration [15]. This study demonstrates that nanosilica incorporation not only accelerated C3S hydration but also led to the formation of more Tobermorite-like structures. Further, irrespective of the temperature effect, the Tobermorite-like structures which were observed in the early stage of hydration changed to Jennite-like structures during the hydration process. The resonances around 909 $cm^{-1}$ and 1076 $cm^{-1}$ due to Si-O stretching vibrations in Jennite-like structures didn't show any narrowing of bands in the nanosilica incorporated samples which indicates C-S-H had a lesser proportion of Jennite-like structure than the samples devoid of nanosilica [16] [17]. Several studies have been carried out demonstrating the accelerating effect of nanosilica in C3S hydration using MIR spectroscopy. However, clearly tracking the formation of C-S-H and CH is challenging in the MIR frequency region due to the fact that resonances in the MIR region are primarily bond specific. Hence, it becomes difficult to differentiate between the resonances due to cement constituents before hydration and those originating due to hydration products. Analysis in the THz frequency region could potentially provide greater insight into the acceleration in hydration due to nanosilica incorporation.

In this work, THz spectroscopy of C3S hydration has been carried out and compared with the corresponding variation in hydration process due to the addition of 2%, 5% and 8% nanosilica. As compared to resonances in the MIR region, which are bond specific, resonances in the THz region are highly sensitive to the crystalline arrangement. Hence, THz spectroscopy can potentially be more effective in analyzing the different hydration products. Since C3S starts to react almost immediately after adding water, experiments have been carried out at an interval of 4 hours till 30 hours and on each day thereafter till 30 days. The variation in the hydration rates has been analyzed by tracking the change in peak intensities as well as a shift in the key resonances. The variation in morphology due to nanosilica incorporation has also been studied by carrying out SEM analysis of the samples.

## 2 Analysis of Variation in Hydration of C3S due to Nanosilica incorporation using THz spectroscopy

### 2.1 Experimental Procedure

Synthetic C3S, purchased from M/s. Sarl Mineral Research Processing, France has been used for carrying out the THz spectroscopy experiments. The nanosilica used in this study is purchased from Nanostructured and Amorphous Materials, Inc, and has an average particle size of 80 nm and a purity of 99 %. Samples of pure C3S and C3S mixed with 2%, 5% and 8% nanosilica have been prepared and stored in separate molds by mixing with water with a water-to-binder ratio of 0.6. Smooth pellets of approximately 1 mm thickness and 13 mm diameter have been prepared by mixing the cementitious materials with High-Density Polyethylene (HDPE) in a ratio of 20:80 by weight. The spectra have been recorded in a Nicolet 6700 Fourier Transform Infrared (FTIR) spectrometer. The spectral analysis was performed in the range of 50 $cm^{-1}$ – 700 $cm^{-1}$ with a spectral resolution of 1.964 $cm^{-1}$. For each pellet, 4 measurements were taken in the transmission mode at different points to average out the inhomogeneity of the sample in the matrix and minor variation in thickness. The average of all the measurements has been considered for the analysis.

### 2.2 Comparison of Variation in Early-Stage Hydration of Nanosilica Incorporated C3S

In this work, the variation in the rates of hydration due to nanosilica incorporation has been analyzed using THz spectroscopy. For the analysis of the data, the plots have been normalized and then baseline corrected. To study the hydration kinetics of C3S, early-stage reactions have been investigated in greater detail (at an interval of 4 hours during the initial 30 hours of hydration). As shown in fig 1, the key resonances observed in this frequency region have been divided into 4 regions and the characteristics in terms of varying intensity and shift in resonances have been analyzed. A comparison of the THz spectra at the onset of reactions (fig. 1a) shows almost the same peak characteristics for the control sample and the nanosilica incorporated samples. The resonance around 520/514 $cm^{-1}$ (band I in figure) is due to a combination of symmetric and asymmetric bending modes of isolated $SiO_4$ tetrahedra present in the C3S (attributed to the vibrational modes of pure C3S) [18]. The resonance around 451 $cm^{-1}$ (band II) is due to a combination of symmetric and asymmetric bending modes of $SiO_4$ tetrahedra along with some contribution from CaO vibrational modes [18]. The vibrational modes in regions III and IV are primarily due to vibrational modes of CaO polyhedra units with some contribution from $SiO_4$ modes. A comparison of the THz spectra at the 4th hour of hydration (fig. 1b) shows a significant reduction in the intensity of the resonance around 520/514 $cm^{-1}$ in the case of 8% nanosilica incorporated C3S sample. This indicates an early onset of reaction for the nanosilica incorporated samples along with more consumption of C3S. The intensity of the resonance around 451 $cm^{-1}$ does not show any significant variation in intensity even in nanosilica incorporated samples. The band of resonances in region III, around 387 $cm^{-1}$, 361 $cm^{-1}$, 347 $cm^{-1}$, 314 $cm^{-1}$ are observed to decrease in intensity, which is more evident for the nanosilica incorporated samples. A slight reduction in the peak intensity of the resonance around 243 $cm^{-1}$ (region IV) is seen for the 8% nanosilica incorporated sample. A small resonance around 181 $cm^{-1}$ is also observed in this region. Figs. 1c and 1d show the comparison of the THz spectra on the $8^{th}$ and $12^{th}$ hour of hydration which clearly show the varying rates of hydration in the control sample (sample without nanosilica) and the nanosilica incorporated samples. The band of resonances in region III is seen to reduce further by hour 12 as well as the resonance around 243 cm$^{-1}$ which is prominent in the case of the control sample. By hour 20 (fig. 1f), the resonance around 520/514 cm$^{-1}$ for the 5% and 8% nanosilica samples are found to get reduced more which is indicative of the faster reaction. The same characteristics of the THz spectra can be observed at hour 24 and hour 30 (figs. 1g and 1f respectively). However, no significant variation in the intensity of the resonance around 451 cm$^{-1}$ is seen during the early stage of hydration (till 30 hours). This could be explained by the fact that the resonances due to deformations of $SiO_4$ bending modes of C-S-H are also observed around this frequency region which makes it very hard to differentiate between the origins of the modes [5], [18].

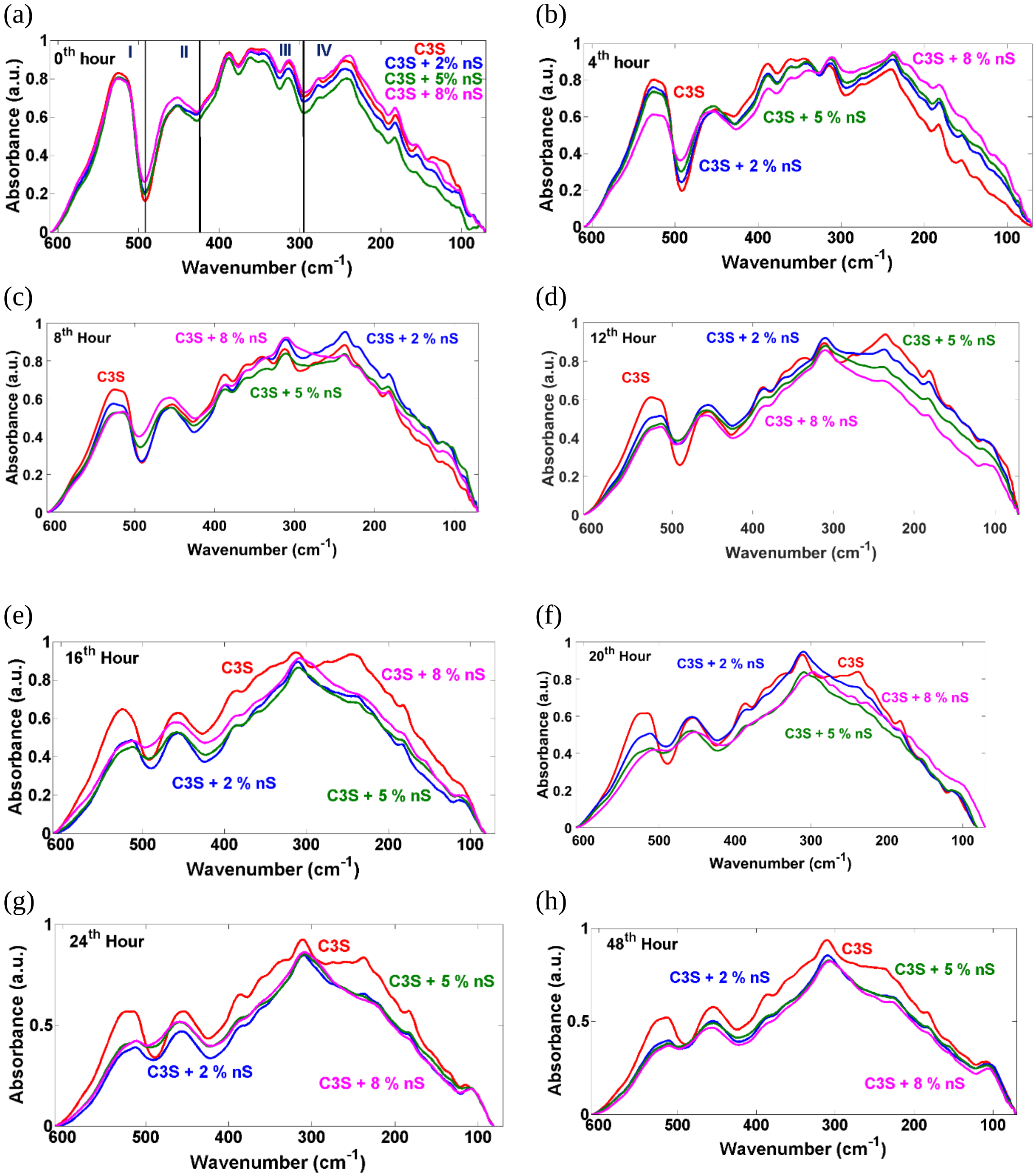

Fig. 1. Comparison of THz spectra to understand the effect of varying percentage of nanosilica incorporated in the C3S during the early stage of hydration (till 48 hours). Variation in the resonance characteristics clearly shows the effect of nanosilica in the hydration rate of C3S.

## 3. Vibrational Frequencies for Silicate Structures with Different Connectivity

As demonstrated from the THz experiments, the characteristics of the resonances related to Si-O bending modes at 451 $cm^{-1}$ and 520/514 $cm^{-1}$ in terms of variation in intensities and shift in wavenumbers are different after the onset of hydration. Formation of C-S-H can be tracked by analysing the shift to higher wavenumbers for the resonance around 451 $cm^{-1}$ which is indicative of C-S-H formation. In the C3S structure, $SiO_4$ monomers are evident which forms into $SiO_4$ chains due to the polymerization effect as C-S-H is formed. This polymerization effect results in the shift to higher wavenumbers. To demonstrate the effect of varying connectivity in silicates contributing to the change in vibrational frequencies, Density Functional Theory (DFT) simulations have been carried out of silicate structures as shown in fig. 2. The structures display geometry optimized structures for silicates with varying connectivity, similar to those found in the layered structure of C–S–H. The quantum chemical model simulations have been employed on the silicates using Gaussian09 software package using B3LYP theory and 6-31G (d,p) basis set. The energy convergence criteria for both optimization and frequency calculation is set to $10^{-8}$ Hartree.

From the simulations of $SiO_4$ monomers ($Q_0$ units), it is seen that a resonance is seen around 414 $cm^{-1}$ which roughly matches to the experimental resonance around 451 $cm^{-1}$. The origin of this resonance is seen to be due to asymmetric bending of $SiO_4$ tetrahedra. Previously, the authors had carried out a detailed analysis of the variation in key resonances during the hydration of C3S [18]. In the crystal structure simulations of C3S, both symmetric as well as asymmetric bending modes of the different $SiO_4$ monomers were observed in the $SiO_4$ unit cell. There is a shift between the experimental resonances from what has been observed in simulations which may be due to the approximations used in DFT calculations. Apart from the approximations in the DFT calculations, the choice of basis sets, cut-off energy (which are also based on approximations) can often lead to some mismatch between the experiments and simulations. After the onset of hydration, the $SiO_4$ monomers are seen to form $SiO_4$ chains due to the polymerization effect. At the early stage of hydration, dimers ($Q_1$ units) are observed in C-S-H. Simulations of the dimers show resonances at 412 $cm^{-1}$ and 433 $cm^{-1}$. The resonances at these frequencies can be attributed to deformations of $SiO_4$ chains. Simulation of C-S-H like structures at a more advanced stage of hydration, such as $SiO_4$ trimers ($Q_2$) show resonances around 412 $cm^{-1}$ and 440 $cm^{-1}$. The resonances at these frequencies can also be attributed to the deformations of $SiO_4$. Simulation of $Q_3$ and ring structure units show prominent resonances getting more prominent around the 450 $cm^{-1}$ (assigned to similar deformations of $SiO_4$ chains). The simulations demonstrate what has been observed in the experiments. With the increase in the chain lengths which indicate a greater degree of polymerization, there is a shift in wavenumbers to higher values in case of the resonance around 450 $cm^{-1}$. The results obtained confirm that Si–O vibrational frequencies indeed increase with increasing degree of connectivity, and the strains in ring structures contribute further to the up-shift of Si–O–Si stretching vibrational frequencies. In our earlier work [18], the variation in resonances during the 28 days of hydration of C3S has been explained by carrying out the crystal structure simulations of C3S (for the anhydrous C3S) and Tobermorite polymorphs, Jennite and Portlandite for the hydrated C3S in CASTEP – Materials Studio software package. Comparison of the vibrational modes of the polymerized chains in crystal structure simulations of C-S-H to those of the simpler models as carried out in this work show a similar deformation in $SiO_4$.

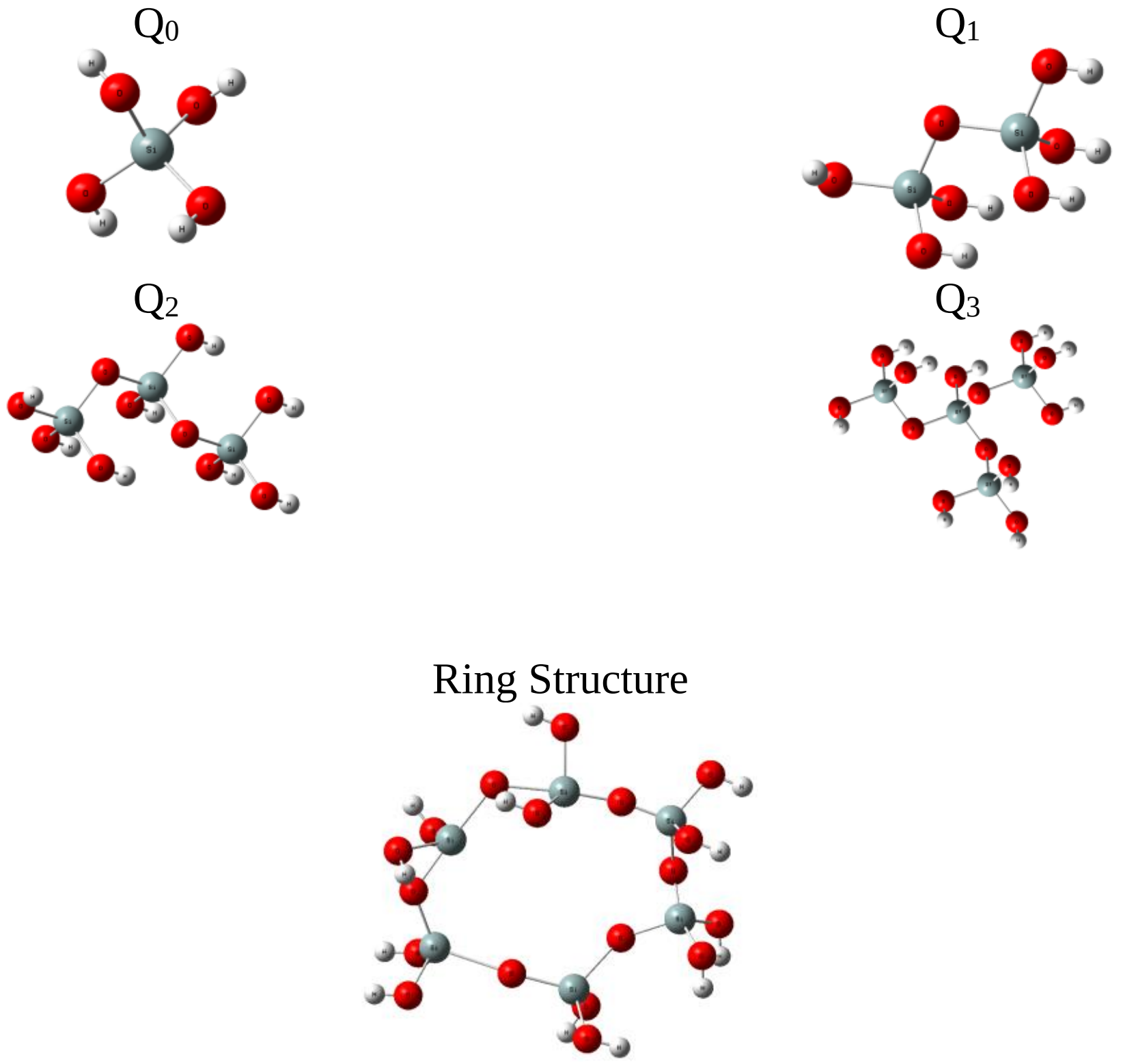

Fig 2. Silicate structures with varying connectivity

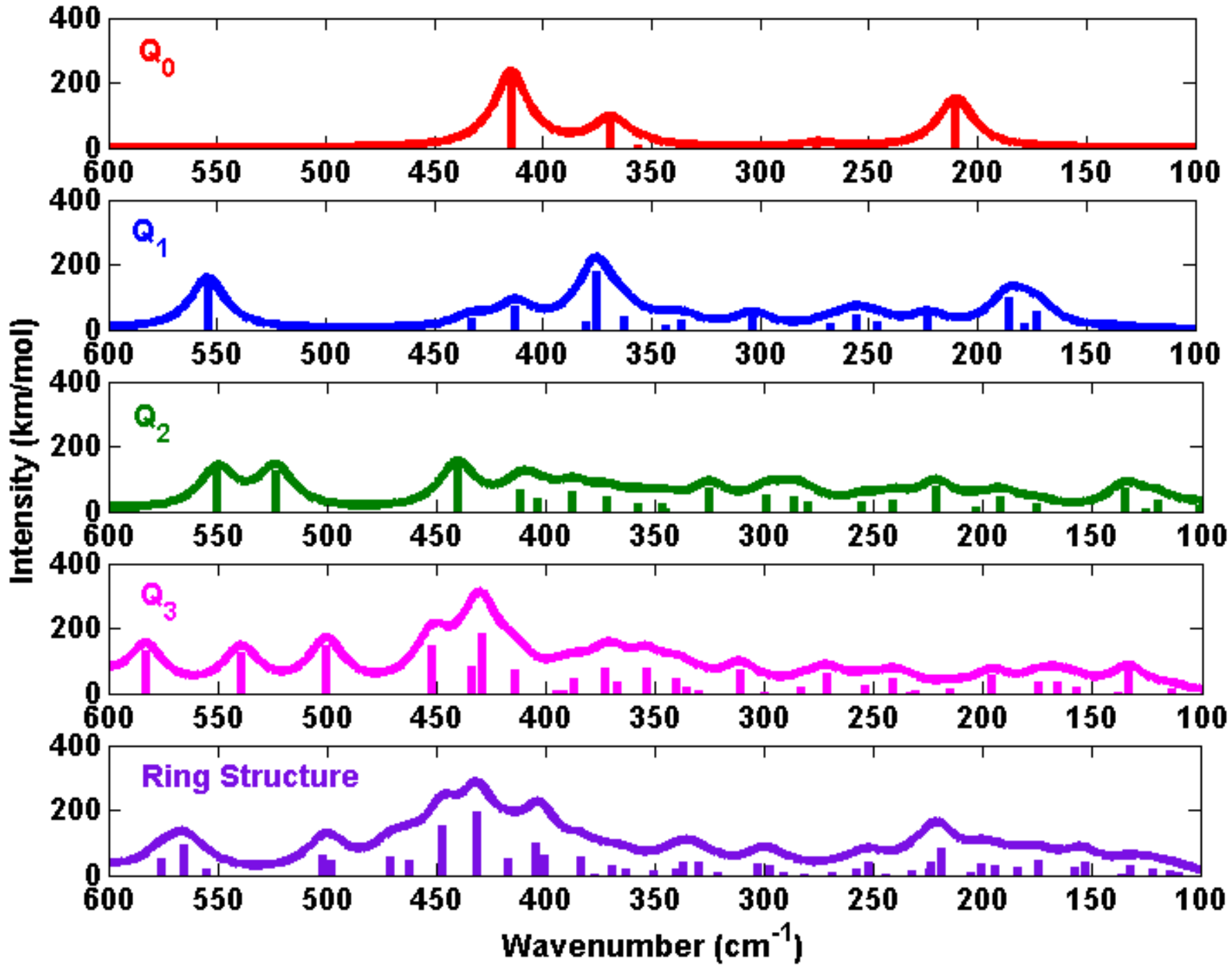

Fig. 3. Comparison of the variation in the THz spectra as a result of the increasing connectivity of the silicates due to polymerization as seen during the formation of C-S-H

### Tracking the Reaction of Nanosilica Incorporated C3S over a period of 30 days

The hydration of nanosilica incorporated C3S has been tracked over a longer term i.e., 30 days, as shown in fig. 4.

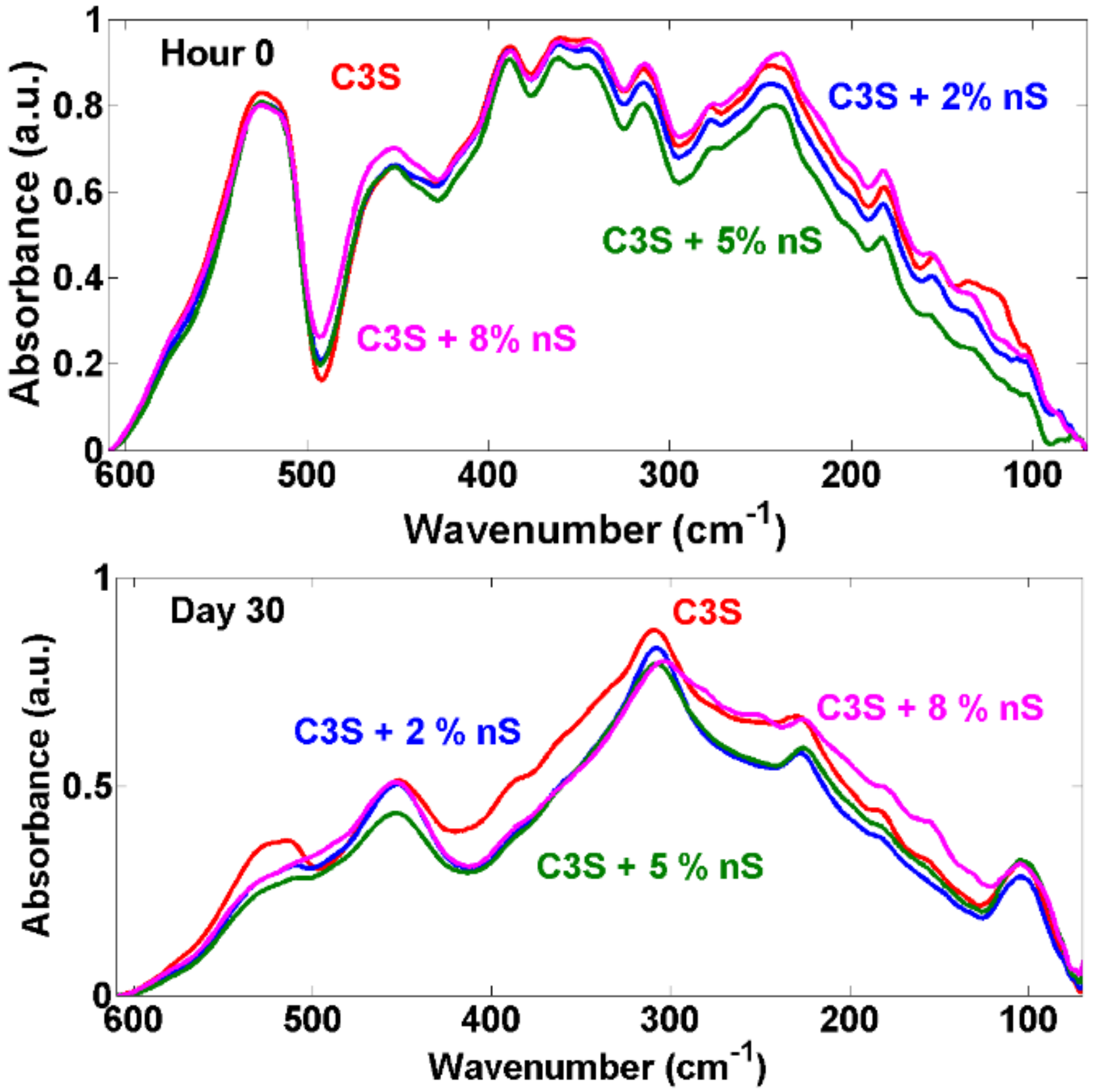

Fig 4. Comparison of the THz spectra of C3S at the onset of reactions and after 30 days of hydration

By tracking the reduction in the peak area of the resonance around 520/514 $cm^{-1}$ (region I) of all the samples, the rate of reaction of C3S during the hydration process can be quantified (as seen in fig. 5). A comparison of the area shows a reduction in the intensity of the nanosilica incorporated C3S samples from a very early stage and at a faster rate as evident from the steeper slope of the spectra from the nanosilica incorporated samples. It has been observed that the peak is reduced till 24 hours after which the area is seen to increase for all the samples till day 4 after which the area is again seen to decrease. After around 7 days, no prominent variation in the area of the resonance is observed until 30 days, which could indicate that the majority of the C3S hydration has been finished.

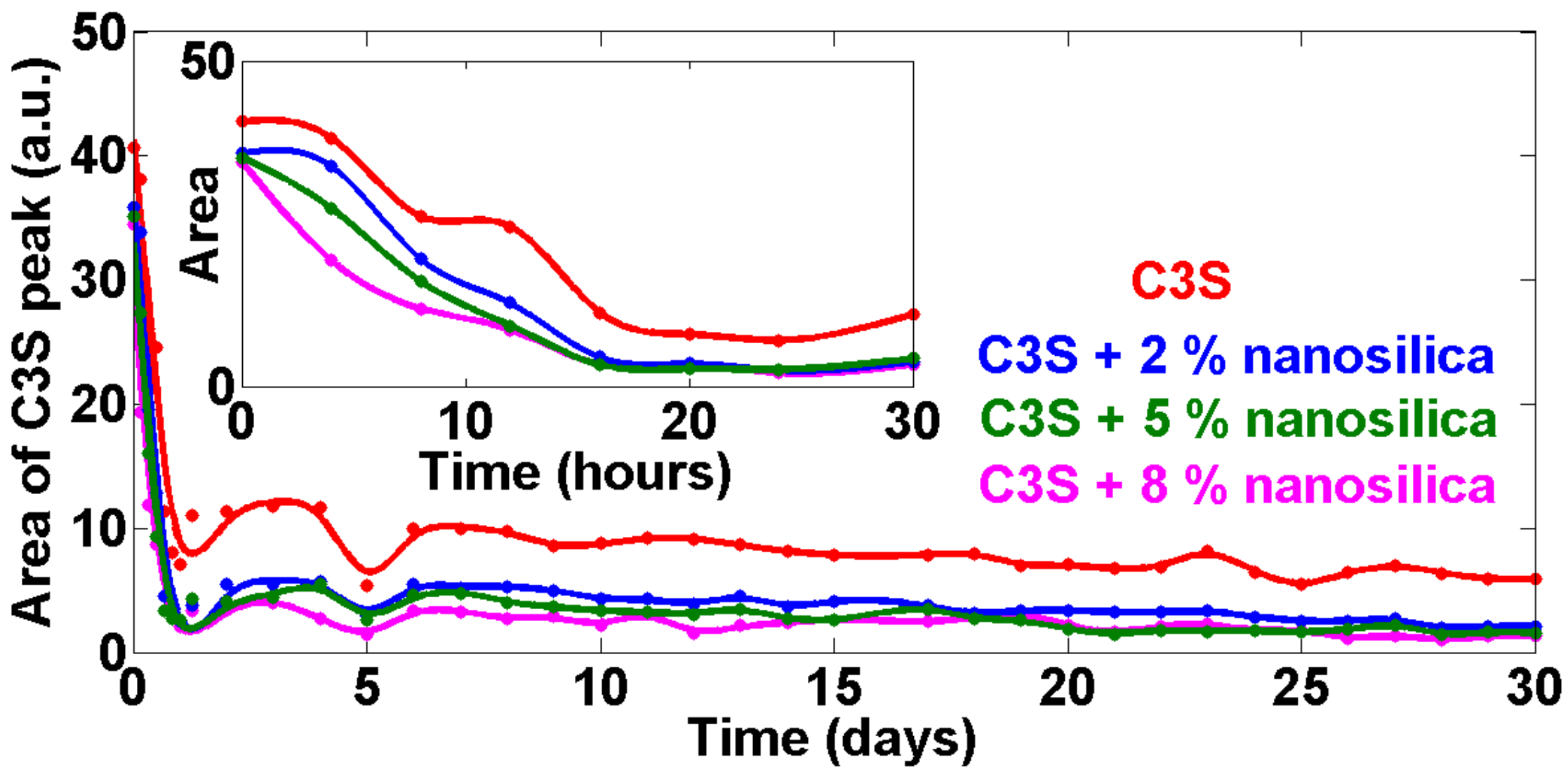

Fig 5. Tracking the rate of C3S reaction for the control and nanosilica incorporated sample by calculating the area corresponding to 520/514 $cm^{-1}$ resonance. The comparison shows that the rate of hydration of C3S for the nanosilica incorporated samples is significantly higher than the control sample. The inset figure shows the accelerating effect of nanosilica incorporation in the hydration of C3S in the very early stage of the process

Density Functional Theory (DFT) simulations of Portlandite like structures by the authors and earlier work of IR spectroscopy of Portlandite which is similar to CH like structures in cementitious systems show a prominent narrowband resonance around 300 $cm^{-1}$ (primarily due to vibrational motion of -OH groups) along with a less intense band near 380 $cm^{-1}$ [16], [18]. As discussed in equation (1), the CH that is formed during hydration reacts with nanosilica in the matrix of C3S to produce more C-S-H. From this, it can be implied that the nanosilica incorporated samples should have a lesser quantity of CH compared to the control sample. This can be concluded by quantification of the peak area of the resonance around 300 $cm^{-1}$ (as shown in fig. 6). Since the analysis has been done by normalizing the resonance around 300 $cm^{-1}$, the reduction in peak area implies further sharpening of the resonance. This confirms the formation of more ordered structures in the hydration products of nanosilica incorporated samples. Compared to the control sample, the nanosilica incorporated samples show a significant decrease of the CH area confirming that the hydration product has reacted with the nanosilica in the C3S matrix. Hence, these results imply that more C-S-H has been formed during hydration.

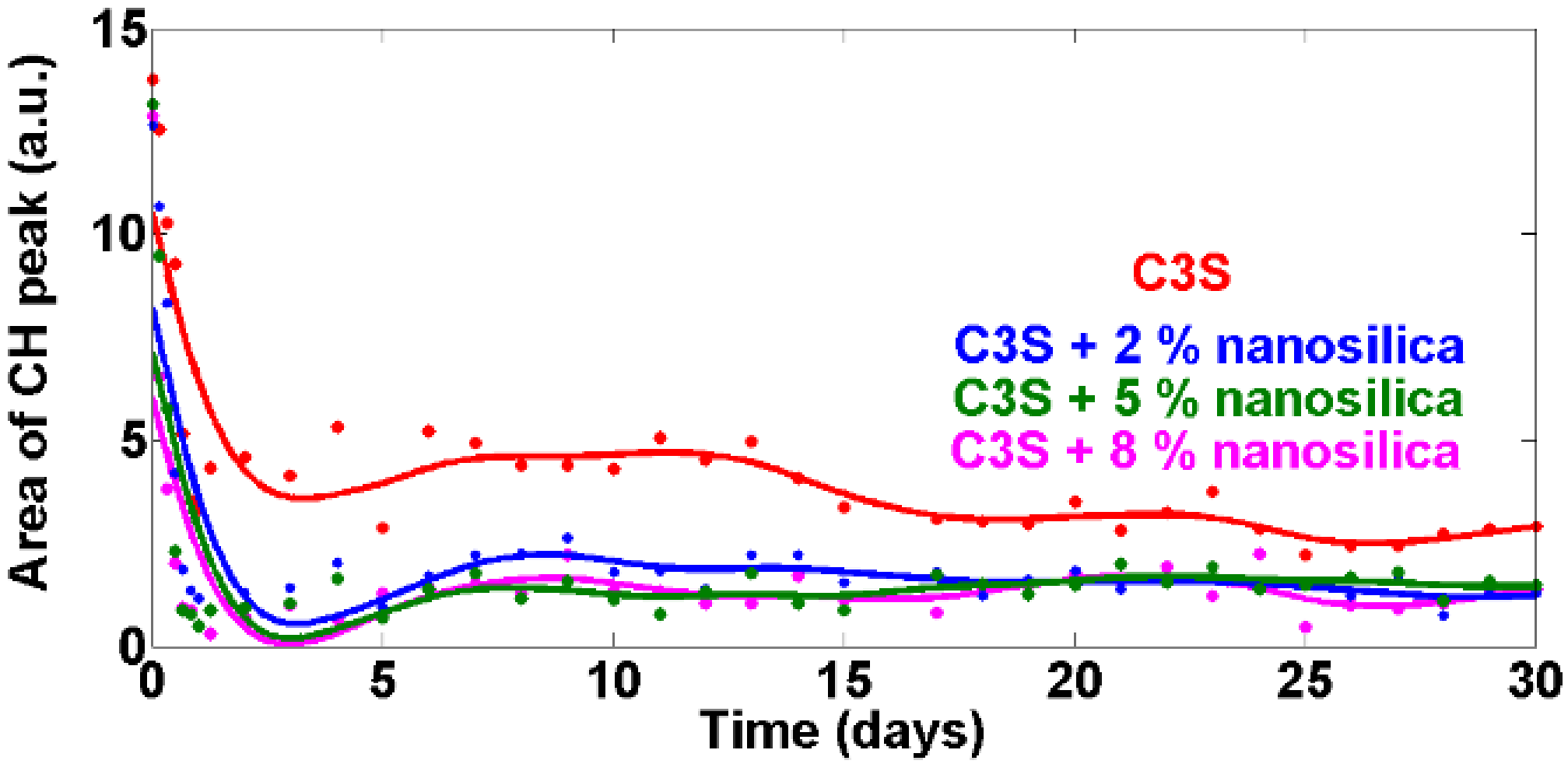

Fig 6. Tracking the area of the resonance around 300 $cm^{-1}$ demonstrates the consumption of CH during hydration. As CH reacts with the nanosilica in the C3S to form C-S-H, the area of the nanosilica incorporated sample shows significantly less area compared to the control sample

### 2.4 Comparison of Hydration Kinetics due to Nanosilica incorporation in C3S using THz spectroscopy by tracking the Shift in Resonances

The rate of hydration of C3S due to nanosilica incorporation can also be analyzed in terms of a shift in key resonances. Earlier studies using IR spectroscopy to track polymerization in silicates indicate the type of C-S-H gel formed [16], [19]. A comparative study of the hydration kinetics of cement with C3S (as major constituent) by the authors have shown that as the hydration progresses, the resonances in this frequency region show varying degree of shift in resonances [18]. Here, the shift in resonances due to the incorporation of nanosilica for the resonances around 514 $cm^{-1}$, 451 $cm^{-1}$ and 314 $cm^{-1}$ has been tracked.

**Tracking the peak shift of the 514 $cm^{-1}$ resonance:** The resonance in this frequency region is mainly due to symmetric and asymmetric bending of $SiO_4$ monomers. As the hydration progresses, the resonance is shifted to lower wavenumbers (fig. 7), to around 513 $cm^{-1}$ for the control sample within the first 3-4 days of hydration after which no significant shift of wavenumber is observed till the period of 30 days. For the nanosilica incorporated samples, the shift in wavenumber is seen to occur earlier (as shown in fig. 7) and to 509 cm$^{-1}$ which is more as compared to the control sample. In the case of the C3S sample with 8% nanosilica, the resonance shifts to around 511 cm$^{-1}$ within the first day of hydration and around 509 cm$^{-1}$ after 5 days of hydration. Similarly, for the other nanosilica incorporated samples (2% and 5%), a similar trend of the shift to lower wavenumbers is seen, although it is slightly delayed as compared to the 8% nanosilica incorporated C3S.

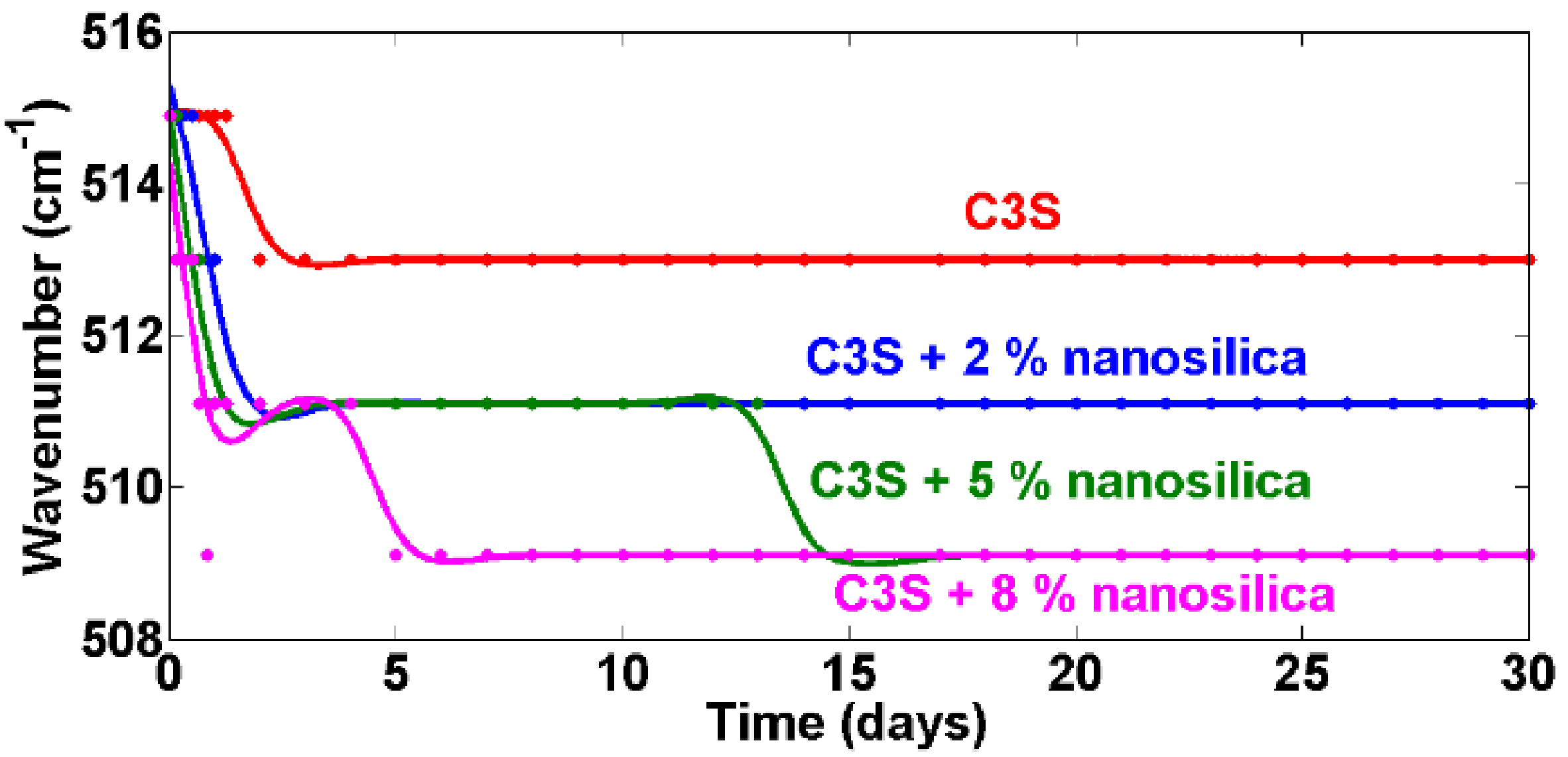

Fig 7. Tracking the peak shift of the resonance around 514 cm$^{-1}$. The comparison shows that the shift to lower wavenumbers for the nanosilica incorporated samples as compared to the control sample is more and starts at an early stage of hydration

**Tracking the peak shift of the 451 cm$^{-1}$ resonance:** The resonance around 451 cm$^{-1}$ is attributed to symmetrical and asymmetrical bending modes of $SiO_4$ monomers with partial contribution from CaO polyhedra. As the hydration progresses, the resonances are shifted to higher wavenumbers (fig. 8). This is considered to be fingerprint evidence of the polymerization effect due to the formation of C-S-H during hydration. As also observed in the resonance around 514 cm$^{-1}$, the shift in wavenumbers is seen to start at an earlier stage of hydration for the nanosilica incorporated samples. In the 8% nanosilica incorporated C3S, the peak wavenumber is shifted to around 464 cm$^{-1}$ from 451 cm$^{-1}$ during the first day while the control sample only shows a shift to around 455 cm$^{-1}$ which implies a lesser degree of polymerization. The samples with 2% and 5% nanosilica show more shift as compared to the control sample but lesser than the 8% nanosilica incorporated sample. The work carried out by Yu et al. has mentioned that the polymerization effect leads to a shift in wavenumbers to higher values, which confirms the formation of Tobermorite like structures [16].

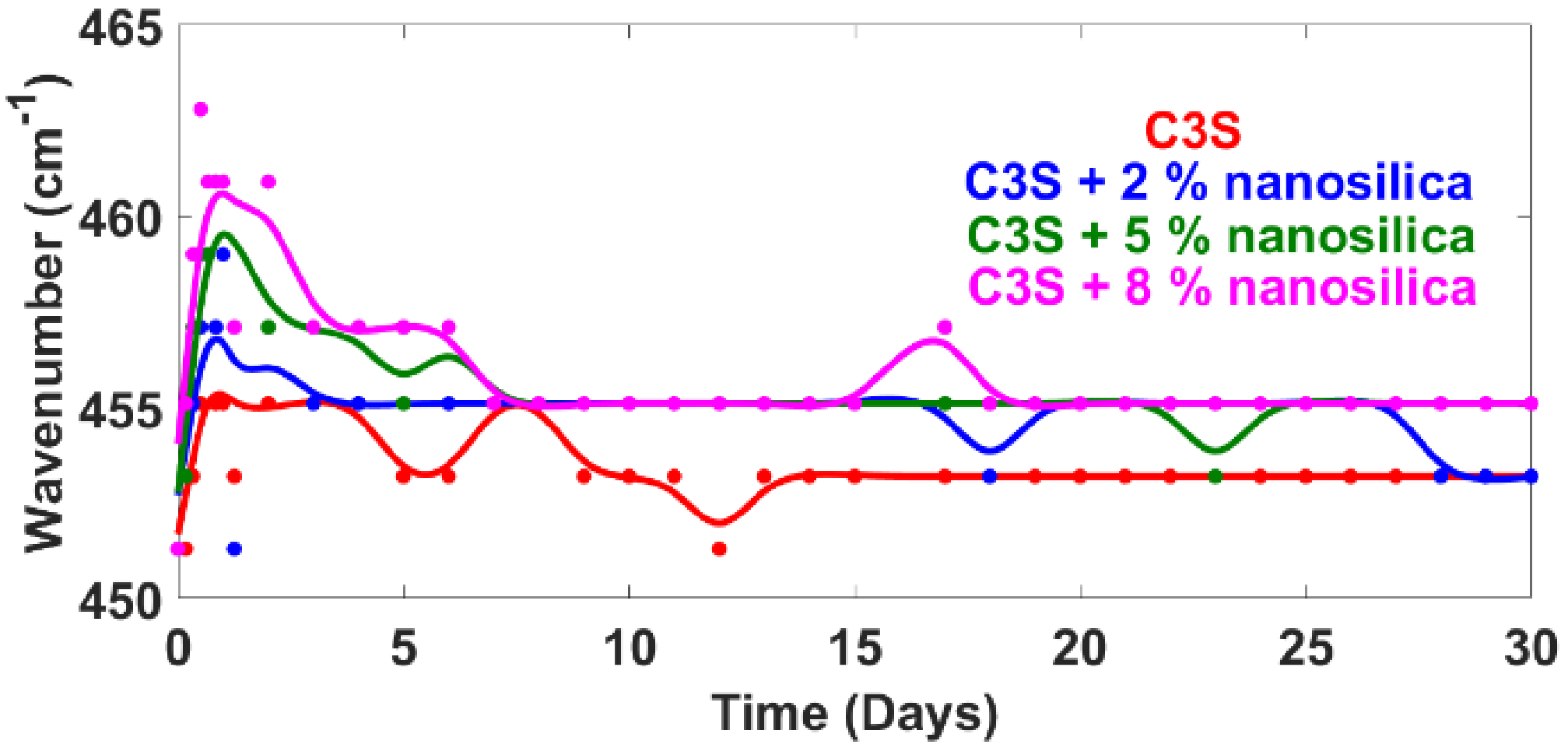

Fig 8. Tracking the peak shift of resonance around 451 $cm^{-1}$. The analysis shows the shift is to higher wavenumbers due to the polymerization of the silicates during the formation of C-S-H. Shift to higher wavenumbers could indicate the formation of Tobermorite like structures.

**Tracking the peak shift of the 314 $cm^{-1}$ resonance:** This resonance around 314 $cm^{-1}$ seems to get sharper as the hydration proceeds along with a shift to lower wavenumbers, around 309 $cm^{-1}$ for the control sample and 305 $cm^{-1}$ for the 8% nanosilica incorporated C3S sample (fig. 9). IR spectroscopy in the far-infrared region (FIR) have shown that portlandite has a prominent resonance around 300 $cm^{-1}$ primarily due to the vibrational motion of the -OH groups along with some contribution from CaO. Synthesized C-S-H with a high C/S ratio (~ 1.70) shows a prominent resonance around 300 $cm^{-1}$ which indicates the existence of portlandite in such structures [16]. A study of C-S-H blended cement using infrared spectroscopy has mentioned that the shift in resonances to lower wavenumbers can be attributed to the transformation of Tobermorite-like structures to Jennite-like structures which also indicates towards the formation of a less compact C-S-H [19].

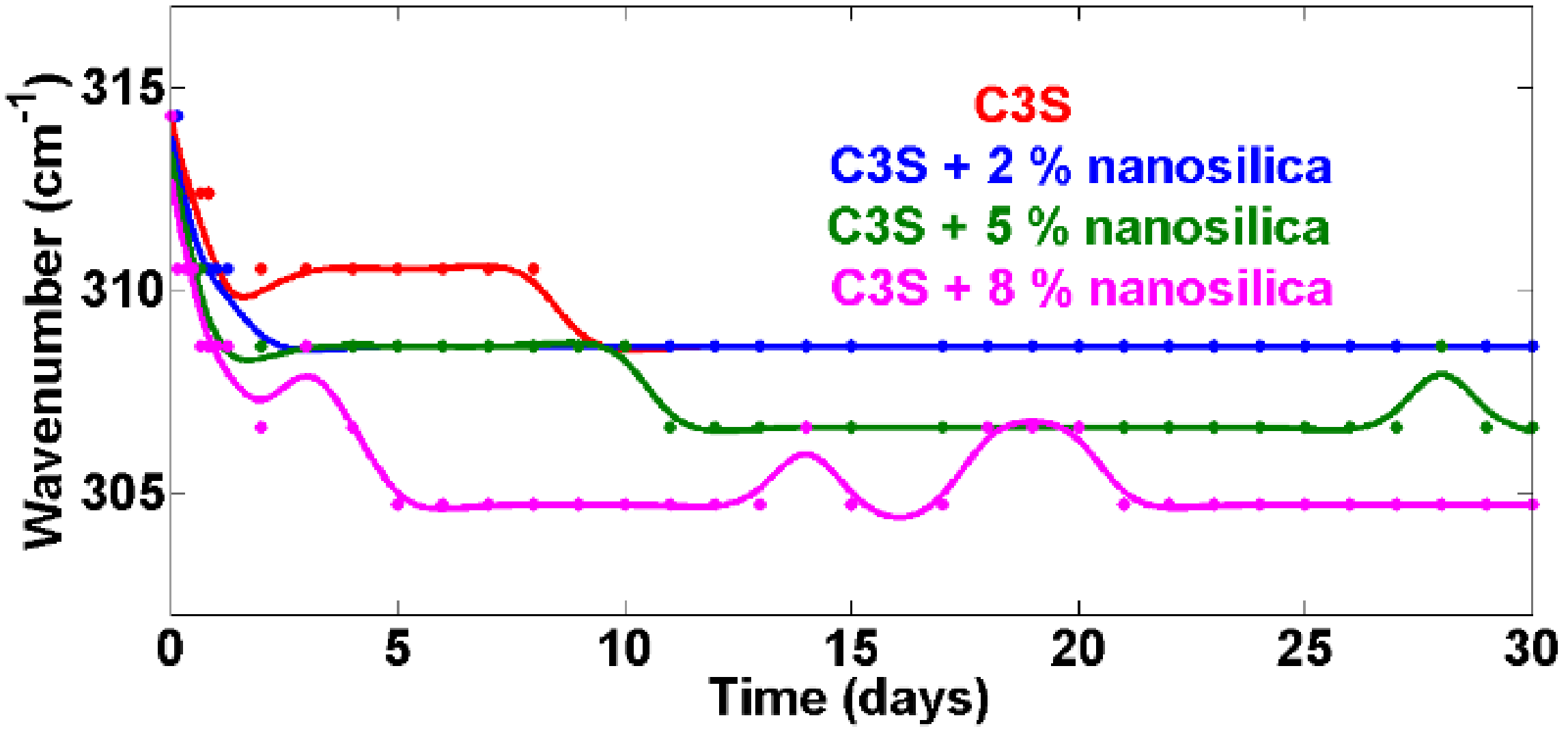

Fig 9. Tracking the peak shift of the resonance around 314 $cm^{-1}$. The comparison shows the larger and earlier shift of resonance to lower wavenumber for the nanosilica incorporated sample.

## 4 Comparison of the Variation in Morphology of in Nanosilica Incorporated C3S using Scanning Electron Microscopy

### 4.1 Experimental Procedure

Scanning Electron Microscopy (SEM) of C3S has been carried out after arresting the hydration of the samples by storing the samples in acetone. Before carrying out the experiments, the samples were taken out from acetone and air-dried for approximately 6 hours to remove any moisture content. FEI Quanta 200 HR-SEM instrument has been employed for the SEM measurements. The samples were non-conductive, which can cause a large accumulation of charge at the sample surface during imaging. To suppress this phenomenon, the samples were coated with a thin gold film.

### 4.2 Analysis of Morphological Variation

A comparison of the morphologies of C3S at the onset of hydration has been compared with the morphologies after 28 days and the variation in morphologies due to nanosilica incorporation in the C3S matrix has been analyzed in detail. Fig. 10 shows the comparative study of 2% and 5% nanosilica incorporated C3S from the onset of reactions to 28 days. During the onset of reaction (0th hour), the presence of nanoparticles is evident in the nanosilica incorporated samples – which is more in the case of the 5% nanosilica C3S matrix. By day 1, all three samples show prominent fiber-like structures that indicate the formation of C-S-H. A comparison of the morphologies on day 7 shows a significant change in the microstructure of the 5% nanosilica incorporated C3S. Even though the fiber-like structures persist for the control sample and 2% nanosilica incorporated samples, there is an agglomeration of the fiber in the case of the 5% nanosilica incorporated sample. Day 28 shows a significant change in the microstructures. A slight variation in the fiber-like structures for the control sample is seen while the nanosilica incorporated samples show much denser morphology. This implied the formation of an advanced stage of C-S-H in the hydration process due to the nanosilica incorporation.

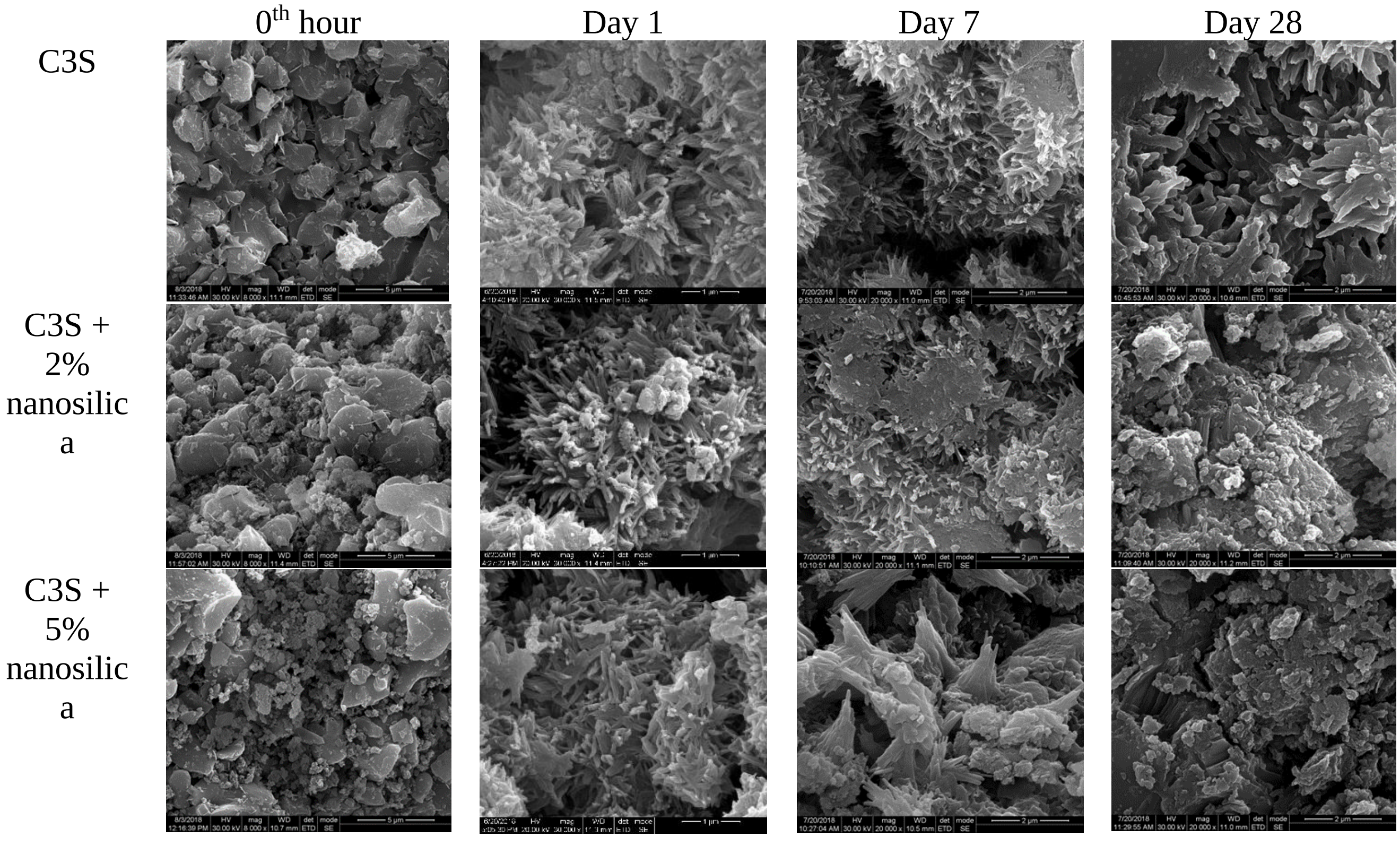

Fig 10. Comparison of the morphologies of the nanosilica incorporated C3S on the 0th hour, day 1, day 7 and day 28. Significant variation in morphologies seen for the samples as hydration progresses. Fiber-like structures that are indicative of the formation of C-S-H can be seen as early as day 1. Day 7 also shows the existence of such fiber-like structures but the nanosilica incorporated samples show a denser morphology which is more prominently evident in day 28 indicating accelerated formation of C-S-H.

## 5 Conclusion

Terahertz spectroscopy of nanosilica incorporated C3S has been studied and the variation of hydration kinetics has been explained in terms of variation in the intensities and shift of resonances. From the variation in intensities, the faster rate of hydration for the nanosilica incorporated samples can be concluded, as evident in the case of the resonance around 520/514 $cm^{-1}$ (due to pure C3S) for the nanosilica incorporated samples. However, no prominent variation in intensity is seen in the case of the resonance around 451 $cm^{-1}$ even after hydration. The origin of this resonance is due to pure C3S before hydration whereas during hydration vibrational modes corresponding to the deformations of $SiO_4$ polymerized chains in C-S-H contribute to significant intensity. In terms of a shift in wavenumbers, the resonance around 510 $cm^{-1}$ and 314 $cm^{-1}$ is seen to shift to lower wavenumbers during hydration. However, for the resonance around 451 $cm^{-1}$, which shifts to higher wavenumbers, this is explained to be due to the polymerization of silicates. Scanning Electron Microscopy (SEM) analysis shows the formation of fiber-like structures for all the samples by day 1. For the 5% nanosilica incorporated C3S, there is an agglomeration of the fibers by day 7. By day 28, a slight variation of the fiber-like morphologies is observed for the control sample while the nanosilica incorporated sample shows signification variation in the morphologies as a denser matrix is observed. Thus, in this work, THz spectroscopy and Scanning Electron Microscopy techniques have been successfully employed to study and clearly demonstrate the accelerating effects of nanosilica incorporation in the C3S hydration.

**Acknowledgements** Shaumik Ray would like to thank CSIR – Senior Research Fellowship for the financial support.